\documentclass[12pt,preprint]{aastex} 

\usepackage{natbib}
\usepackage{epstopdf}
\usepackage{graphicx}
\usepackage{subfigure}
\usepackage{rotating}
\usepackage{textcomp,graphicx,amsmath,pdflscape,amssymb}

\begin{document}

\def\Msun{M_\odot}

\def\Lsun{L_\odot}

\def\Rsun{R_\odot}

\slugcomment{Submitted to ApJ}

\shorttitle{Nova white dwarf masses}

\shortauthors{Shara et al}

\title{The Masses and Accretion Rates of White Dwarfs\\ in Classical and Recurrent Novae}

\author{Michael~M.~Shara\altaffilmark{1,2}, Dina Prialnik\altaffilmark{3}, Yael Hillman\altaffilmark{1,4} and Attay Kovetz\altaffilmark{5}}


\altaffiltext{1}{Department of Astrophysics, American Museum of Natural History, Central Park West and 79th Street, New York, NY 10024-5192, USA}

\altaffiltext{2}{Institute of Astronomy, University of Cambridge, Madingley Road, Cambridge CB3 0HA, United Kingdom}

\altaffiltext{3}{Department of Geosciences, Tel Aviv University, Ramat Aviv, Tel Aviv 69978, Israel}

\altaffiltext{4} {Department of Particle Physics and Astrophysics, Weizmann Institute of Science, 76100 Rehovot, Israel}

\altaffiltext{5} {School of Physics and Astronomy, Faculty of Exact Sciences, Tel-Aviv University, Tel Aviv, Israel}

\begin{abstract}
Models have long predicted that the frequency-averaged masses of white dwarfs in Galactic classical novae are twice as large as those of field white dwarfs. Only a handful of dynamically well-determined nova white dwarf masses have been published, leaving the theoretical predictions poorly tested. The recurrence time distributions and mass accretion rate distributions of novae are even more poorly known. To address these deficiencies, we have combined our extensive simulations of nova eruptions with the \citet{str10} and \citet{sch10} databases of outburst characteristics of Galactic classical and recurrent novae to determine the masses of 92 white dwarfs in novae. We find that the mean mass (frequency averaged mean mass) of 82 Galactic classical novae is 1.06 (1.13)~$M_\odot$, while the mean mass of 10 recurrent novae is 1.31~$M_\odot$. These masses, and the observed nova outburst amplitude and decline time distributions allow us to determine the long-term mass accretion rate distribution of classical novae. Remarkably, that value is just $1.3\times10^{-10}M_\odot$/yr, which is an order of magnitude smaller than that of cataclysmic binaries in the decades before and after classical nova eruptions. This predicts that old novae become low mass transfer rate systems, and hence dwarf novae, for most of the time between nova eruptions. We determine the mass accretion rates of each of the 10 known Galactic RN, finding them to be in the range $10^{-7}$ - $10^{-8}$ $ M_\odot$/yr.  We are able to predict the recurrence time distribution of novae and compare it with the predictions of population synthesis models.

\end{abstract}


\section{Introduction}

Novae are powered by thermonuclear runaways (TNRs) in the
hydrogen-rich envelopes of white dwarf (WD) stars \citep{sta75,pri78}. Those envelopes are accreted from 
brown dwarf, red dwarf, subgiant or red giant companions in the stellar binary systems known as cataclysmic variables (CVs). 
WD envelopes become increasingly electron degenerate as their masses increase. Once a critical pressure is reached, 
the timescale for nuclear energy generation at the base of a WD hydrogen-rich envelope 
becomes smaller than the timescale to transport away that energy. A TNR inevitably 
ensues \citep {sha81}, with the resulting nova reaching maximum magnitude in the range M = -5 to -10.7 \citep{sha09}. 

The degenerate equation of state of a WD ensures that its radius decreases and its gravitational potential greatly increases  
as its mass increases \citep{cha31,cha35}. Thus, with increasing WD
mass, less hydrogen can be accreted onto a WD before a TNR occurs
\citep {sha81}. All other things being equal, lower mass envelopes can be ejected 
faster than those of higher mass, so nova TNRs on massive WDs should eject their envelopes,
and begin to decline in brightness, faster than those on low mass WDs. 
Thus if WD mass was the {\it only} free parameter in nova binaries then novae would be luminous, well-understood 
standard candles displaying negligible scatter \citep{sha81,liv92}. The WD mass would be tightly correlated with, and directly measurable
from the rate at which a nova declines in brightness. 

In fact, nova eruptions display a 3 magnitude scatter about the
so-called Maximum Magnitude - Rate of Decline Relation (MMRD) \citep{pri95,yar05,hac10}, rendering the MMRD nearly useless
as a distance indicator \citep{fer03,kas11,sha17}. Just as important, the MMRD relation cannot be used to determine WD masses.

Several factors, in addition to WD mass, influence each nova outburst \citep{sta75,sha80}. 
These are the accretion rate onto the WD and the resulting envelope mass \citep{pac78,pri82}; 
the WD luminosity \citep{pri95,yar05}; its chemical composition (He, CO or ONe), and the chemical composition of the accreted
matter (H-rich or He) \citep{fau72,kov85,sta86}. For primarily hydrogen-accreting, CO WDs, the accreted mass is 
the next most important parameter after the WD mass in determining a nova's outburst properties \citep{sha17}.

There is compelling evidence that the mass of the WD in a CV is not static in time. The highly CNO or Ne-enriched ejecta of some novae demonstrate that 
WD material is being ablated during some nova eruptions \citep{wil78,gal80,wil82,geh08}. The existence of a WD in nova M31N 2008-12a (hereafter M31-12a) which recurs at least annually \citep{hen15} demands a WD close to the Chandrasekhar mass \citep{kat14}. No trace of neon is seen in ultraviolet spectra of its ejecta \citep{dar17}. The CO WD in RS Oph is nearly as massive ($\sim 1.3 M_\odot$) \citep{mik17}.
Since CO WDs cannot be born with masses larger than $\sim 1.1 M_\odot$ \citep{ibe85,rit96}, 
the underlying WDs in RS Oph and M31-12a have likely experienced significant mass growth since their births. 

The latter results suggest that WDs in some nova systems may approach the Chandrasekhar mass, 
and may produce type Ia supernovae \citep{whe73,can96,han04,mao14}. Understanding the evolution of WD masses in CVs, 
and their resulting connections to SNIa is thus an important, though challenging task.
While the mass distribution of single WDs near the Sun is sharply peaked at $\sim 0.6 M_\odot$ \citep{ber92,kep16}, 
population synthesis models  \citep{rit91,pol96,nel04,che16} predict a much wider range of nova WD masses. \citet{rit91} and \citet{nel04}
find frequency-averaged masses of WDs in nova systems that are 1.14+/-0.10 $M_\odot$ and 0.95+/-0.15 $M_\odot$, respectively. 
They note that these values are in good agreement with the observational estimate of 0.90 $M_\odot$ \citep{rit91}. This latter WD mass estimate 
was based on 8 novae; none of those WD masses were dynamically determined. 

In fact, accurate, dynamically-determined measurements of nova WD masses and frequency-averaged WD masses 
are extremely difficult to make, and few in number, because the WD radial velocities are 
almost always masked by their accretion disks \citep{you80,smi98}. \citet{sio99} has summarised the range of observational data 
and assumptions that goes into determining nova WD masses (including parameters of emission-line profiles, velocity width at half
or mean intensity, rms line widths, and velocity separation of doubled emission peaks), as well as the assumption of Roche geometry 
and an empirical mass-radius relation for the secondary stars. The host of assumptions and approximations, and the small number of nova systems 
for which these procedures have been carried out, suggest  that one of the most direct and critical predictions of 
stellar binary population synthesis models - the mass distribution of WDs in novae - has not been adequately tested. The theoretically-predicted 
mass transfer rate and recurrence time distributions are even less well-determined.

The goal of this paper is to demonstrate a methodology for determining the mass of a WD in a nova CV, utilizing 
only our extensive simulations of novae combined with observables that are straightforward to measure for any nova. 
Extensions of the methodology also yield the mass transfer rate and recurrence time distributions of Galactic novae.
We apply this methodology to determine the masses and mass transfer rates of 82 WDs in classical novae (CN), and another 10 in recurrent novae (RNe). 
We also derive the recurrence time distribution for those same 82 CN WDs. This is by far the largest sample of novae for which these parameters 
have been derived in a uniform and consistent manner. The results enable the strongest tests to date of nova population synthesis models' predictions. 

In Section 2 we summarize our nova models and the relationships between model parameters --- WD mass and accretion rate --- and the resulting nova characteristics, such as nova outburst amplitude,  envelope mass and decline time, which enable the determination of WD masses and accretion rates based on observations. In Section 3 we describe the observations of Galactic RNe and CN which serve as the databases for this study.
In section 4 we use the model-derived relationships to determine the masses of the WDs in the databases, as well as the accretion rates and recurrence times and their distributions. We compare the results with the few dynamically measured nova WD masses, and with the predictions of nova WD masses from population synthesis models. We also compare our WD mass results with those of derived from a completely different methodology - nova atmospheric models. In section 5 we determine and discuss the {\it true} WD mass and mass accretion rate distributions in Galactic novae. Our conclusions are summarized in section 6.

\section{Nova Models: relationships between nova characteristics}

One of the most extensive and widely used ensembles of nova models is that of \citet{yar05}.
White dwarf masses of 0.4, 0.65, 1.00, 1.25 and 1.40 $M_\odot$ and core temperatures of 10, 30 and 50 MKelvins,
were allowed to accrete solar composition matter at rates of $10^{-6}, 10^{-7}, 10^{-8},10^{-9},10^{-10},10^{-11}$, and $10^{-12} M_{\odot}$/yr.
Details of the hydrodynamic Lagrangian stellar evolution code, including the opacities, nuclear reactions network, mass-loss algorithm, 
diffusion and accretion heating, and convective fluxes are described in detail in \citet{pri95} and \citet{yar05}. To these we add the models from \citet{hil16}
which more densely cover the parameter space of WDs accreting at the very high rates ($3\times10^{-8}$ to $6\times10^{-7} M_{\odot}$/yr), which can grow WD masses considerably. 

Each of these models predicts the total mass accreted before a TNR, and the outburst characteristics of each of the resulting novae. In particular, we emphasize that the amplitude in visual magnitudes of the outburst $A$, as well as the duration of the mass-loss phase $t_{m-l}$ are calculated by \citet{yar05}. These two parameters are directly observable for many Galactic novae. They enable us to match models specified by just three parameters - white dwarf mass, core temperature and mass accretion rate - with well observed Galactic novae, regardless of their often poorly-determined distances. Of the three model parameters, the least important one by far (in the sense that results are affected by it to a much lesser extent) is the core temperature. We adopt the intermediate WD core temperature of 30 MKelvins in the \citet{yar05} models, and use just the two independent model parameters, $M_{\rm WD}$ and $\dot M$. The results used are given in Tables 1 and 2.
\begin{table}[ht]
\begin{center}
\caption{Nova outburst amplitudes A derived from models \citep{yar05}}
\smallskip
\begin{tabular}{cccccc}
\tableline\tableline
{} & {$M_{\rm WD}$} & {0.65$M_\odot$} & {1.00$M_\odot$} & {1.25$M_\odot$} & {1.40$M_\odot$}\\
{$\log\dot M$} & & & & &\\
\tableline
-8  &$\mid$ & 9.5 & 9.4 & 9.0 & 8.4\\
-9  &$\mid$ & 12.0 & 12.0 & 11.6 & 11.0\\
-10 &$\mid$ & 16.2 & 14.4 & 14.3 & 13.6\\
-11 &$\mid$ & 15.2 & 15.3 & 15.4 & 15.3\\
\tableline
\end{tabular}
\end{center}
\end{table}
\begin{table}[ht]
\begin{center}
\caption{Mass loss times $t_{m-l}$ (log(days)) derived from models \citep{yar05}}
\smallskip
\begin{tabular}{cccccc}
\tableline\tableline
{} & {$M_{\rm WD}$} & {0.65$M_\odot$} & {1.00$M_\odot$} & {1.25$M_\odot$} & {1.40$M_\odot$}\\
{$\log\dot M$} & & & & &\\
\tableline
-8  &$\mid$ & 3.09 & 2.20 & 1.51 & 0.67\\
-9  &$\mid$ & 2.83 & 2.15 & 1.50 & 0.11\\
-10 &$\mid$ & 2.68 & 2.07 & 1.34 & 0.18\\
-11 &$\mid$ & 1.57 & 1.01 & 0.31 & -0.73\\
\tableline
\end{tabular}
\end{center}
\end{table}

\section{Nova Observations: outburst amplitudes and decline times}

The largest uniform compilation of Galactic nova light curves is that of \citet{str10}. The detailed observational characteristics of 93 novae - including the outburst amplitudes A and times $t_{2}$ to decline by two magnitudes from peak brightness - are summarised from almost 230,000 individual measured magnitudes.
The largest uniform compilation of recurrent nova (RN) light curves is that of \citet{sch10}.

{\it High} accretion rates on massive WDs can accumulate critical envelope masses on timescales as short as
years \citep{yar05}. This is the source of the Recurrent Novae (RNe), which have
massive WDs and inter-eruption intervals of a century or less.  RNe have recently been estimated to comprise 25\% of all novae \citep{pag14}.
Examples include M31-12a and RS Oph.
The extraordinary recurrent nova M31N 2008-12a in the Andromeda galaxy, which erupts every year \citep{hen15} 
and fades by 2 magnitudes in just 1.65 days \citep{dar16} must host a WD with a mass within a few percent of the Chandrasekhar mass.
The mass of the WD in RS Oph must lie in the range $1.3\pm0.10 M_\odot$ \citep{mik17}, while that in TCrB is most likely $\sim$1.2$M_\odot$\citep{bel98}.

In Table 1 we list the novae that will serve as our database, and their outburst amplitudes $A$ and times of decline by two magnitudes $t_2$. 
The distribution of decline times $t_2$ is shown in Fig.1.
\begin{figure}[h]
\includegraphics[scale=0.70]{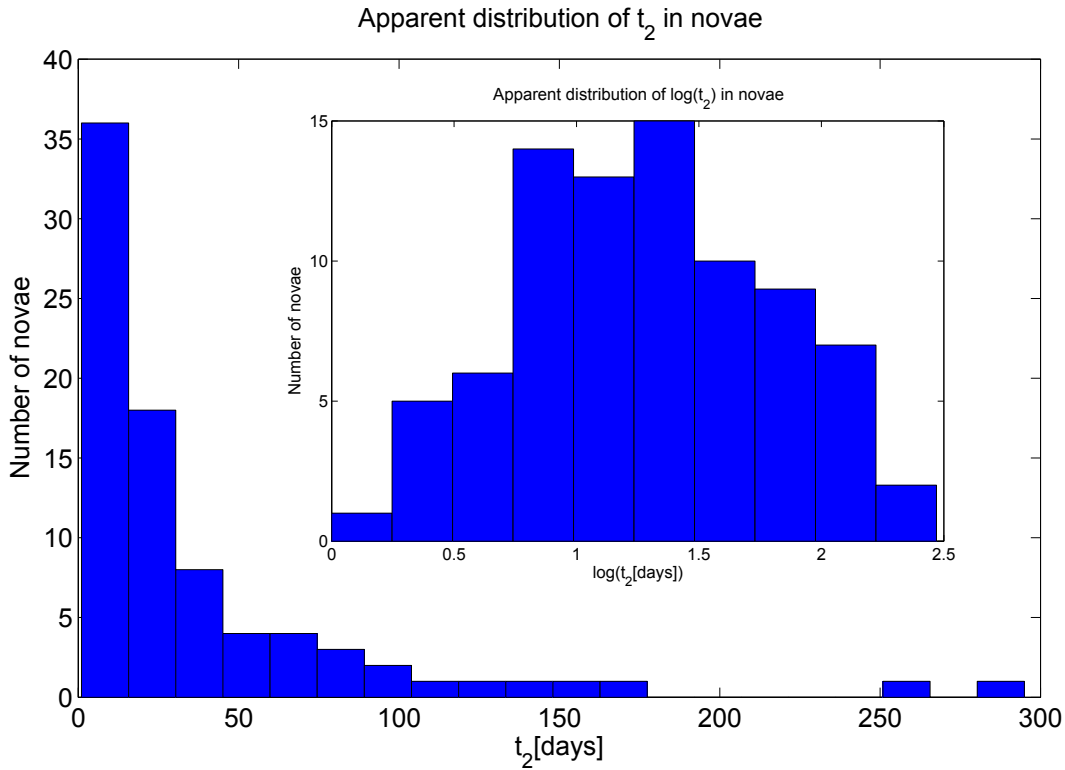}
\caption{The apparent distribution of decline times $t_2$ for the CN sample.}
\end{figure}

\section{The masses of WDs in Galactic CN and RNe and their accretion rates}

As mentioned in Section 2, the grid of models over the $(M_{\rm WD},\dot M)$ space for a given WD core temperature yields, among other characteristics, the outburst amplitudes A and the mass-loss times of novae $t_{m-l}$. Observations yield, among other characteristics, the outburst amplitudes A and the times of decline by two magnitudes $t_2$, as mentioned in Section 3. The working assumption of our study will be that the observationally derived $t_2$ and the model-derived mass loss time $t_{m-l}$ are identical. We denote both by $t_{dec}$,
 \begin{equation*}
 t_2 \equiv t_{dec} \equiv t_{m-l}\ .
 \end{equation*}
Ideally, we would require from observations and models both the duration of the mass-loss phase $t_{m-l}$ and the visual decline time $t_2$. Unfortunately, self-consistent
evolutionary models provide the mass loss phase duration accurately, together with its dependence on the
WD mass and accretion rate, while observations provide accurately the visual decline time. Fortunately, however, these times are closely connected, 
as shown analytically in the Appendix of \citet{sha81}. The maximal luminosity of a nova in the visible
band is obtained when the envelope has become extended enough for mass loss to begin. Near the end of mass loss,
the WD contracts rapidly, the luminosity shifts to shorter wavelengths, and the visual luminosity starts declining. This is
why we have adopted the earlier decline time obtained from observations, that is, $t_2$ rather than 
 $t_3$, $t_4$, etc., as the closest to the mass-loss time derived from modeling.
 
Now, we interpolate in the grid of models given in Tables 1 and 2, using second-order polynomials, to obtain functions $A(M_{\rm WD},\dot M)$ and $t_{dec}(M_{\rm WD},\dot M)$. These relations are then inverted for each pair of observables $A$ and $t_{dec}$ in the sample of \cite{str10}, to determine both the WD mass $M_{\rm WD}(A,t_{dec})$ and accretion rate $\dot M(A,t_{dec})$ of the corresponding nova (a similar procedure was already used by \cite{pri95}). The results are given in Table 3. We omitted three CN from the sample of 85 CN in \citet{str10}: V445 Pup (which is an eruption of a helium envelope, rather than a TNR in a hydrogen-rich envelope), as well as V888 Cen and CP Pup. The amplitude A of V888 Cen is very small, falling near the limits of the $A$ range (similar to those on RNe), while that of CP Pup is very large, so that the WD masses could not be determined.
 
\begin{table}
\begin{center}
\caption{CN: WD masses and accretion rates from observations \citep{str10} and models}
\begin{tabular}{llccccccc}   
\tableline\tableline
\multicolumn{4}{c}{Observations}&&\multicolumn{4}{c}{Modeling}\\
\cline{1-4}\cline{6-9}
&&&&&&&&\\
Year&Nova&A(mag)&$t_2$(days)&&$M_{\rm WD}(M_\odot)$&$\log\dot{M}(M_\odot$/yr)&$p_{rec}$(yr)&$m_{acc}(M_\odot)$\\
\tableline
                 1986&OS And&11.0&11&&1.16&-8.81&6.91E+03&1.07E-05\\
		1925&DO Aql&9.5&295&&0.62&-8.16&1.68E+04&1.16E-04\\
		1936&V356 Aql&11.3&127&&0.85&-8.94&2.85E+04&3.26E-05\\
		1945&V528 Aql&11.6&16&&1.14&-9.07&1.58E+04&1.34E-05\\
		1918&V603 Aql&12.2&5&&1.24&-9.33&1.07E+04&4.97E-06\\
		1970&V1229 Aql&11.5&18&&1.12&-9.03&1.54E+04&1.45E-05\\
		1982&V1370 Aql&10.3&15&&1.13&-8.51&4.14E+03&1.29E-05\\
		1993&V1419 Aql&13.4&25&&1.08&-9.85&1.12E+05&1.57E-05\\
		1995&V1425 Aql&12.0&27&&1.09&-9.24&3.08E+04&1.75E-05\\
		1999&V1493 Aql&10.9&9&&1.18&-8.77&5.38E+03&9.17E-06\\
		1999&V1494 Aql&13.0&8&&1.20&-9.68&3.44E+04&7.22E-06\\
		1891&T Aur&10.4&80&&0.93&-8.55&9.05E+03&2.54E-05\\
		1993&V705 Cas&10.7&33&&1.05&-8.68&9.45E+03&1.97E-05\\
		1995&V723 Cas&8.6&263&&0.77&-7.77&1.51E+03&2.55E-05\\
		1986&V842 Cen&10.9&43&&1.02&-8.77&1.28E+04&2.18E-05\\
		1991&V868 Cen&11.2&31&&1.06&-8.90&1.53E+04&1.93E-05\\
		2001&V1039 Cen&11.7&25&&1.09&-9.11&2.23E+04&1.71E-05\\
		1995&BY Cir&10.5&35&&1.04&-8.60&7.84E+03&1.99E-05\\
		1999&DD Cir&12.6&5&&1.24&-9.50&1.52E+04&4.76E-06\\
		1981&V693 CrA&14&10&&1.15&-10.11&1.19E+05&9.24E-06\\
		1920&V476 Cyg&14.3&6&&1.18&-10.24&1.14E+05&6.53E-06\\
		1970&V1330 Cyg&7.6&161&&0.91&-7.34&2.67E+02&1.22E-05\\
		1975&V1500 Cyg&16.0&2&&1.09&-10.98&6.72E+05&7.09E-06\\
		1978&V1668 Cyg&13.5&11&&1.16&-9.89&7.39E+04&9.42E-06\\
		1986&V1819 Cyg&7.7&95&&0.97&-7.38&2.88E+02&1.19E-05\\
		1992&V1974 Cyg&12.6&19&&1.12&-9.50&4.44E+04&1.39E-05\\
		2001&V2274 Cyg&	8.5&22&&1.13&-7.73&5.09E+02&9.51E-06\\
		2001&V2275 Cyg&11.5&3&&1.27&-9.03&3.37E+03&3.15E-06\\
		2006&V2362 Cyg&12.9&9&&1.19&-9.63&3.45E+04&8.01E-06\\
		2007&V2467 Cyg&11.6&8&&1.20&-9.07&9.53E+03&8.09E-06\\
		2008&V2491 Cyg&12.5&4&&1.26&-9.46&1.11E+04&3.82E-06\\
		1967&HR Del&8.5&167&&0.84&-7.73&1.03E+03&1.92E-05\\
		1912&DN Gem&12&16&&1.14&-9.24&2.29E+04&1.30E-05\\
		\tableline
\end{tabular}
\end{center}
\end{table}
\setcounter{table}{2}
\begin{table}
\begin{center}
\caption{CN: WD masses and accretion rates (\textit{cont.})}
\begin{tabular}{llccccccc}   
\tableline\tableline
\multicolumn{4}{c}{Observations}&&\multicolumn{4}{c}{Modeling}\\
\cline{1-4}\cline{6-9}
&&&&&&&&\\
Year&Nova&A(mag)&$t_2$(days)&&$M_{\rm WD}(M_\odot)$&$\log\dot{M}(M_\odot$/yr)&$p_{rec}$(yr)&$m_{acc}(M_\odot)$\\
\tableline
		1934&DQ Her&12.7&76&&0.95&-9.55&8.87E+04&2.51E-05\\
		1960&V446 Her&11.3&20&&1.11&-8.94&1.36E+04&1.55E-05\\
		1963&V533 Her&12.0&30&&1.07&-9.24&3.24E+04&1.84E-05\\
		1987&V827 Her&10.6&21&&1.10&-8.64&6.92E+03&1.59E-05\\
		1991&V838 Her&13.8&1&&1.35&-10.02&6.44E+03&6.09E-07\\
		1936&CP Lac&13.0&5&&1.24&-9.68&2.21E+04&4.63E-06\\
		1950&DK Lac&7.9&55&&1.03&-7.47&3.38E+02&1.15E-05\\
                 1939&BT Mon&7.6&118&&0.95&-7.34&2.57E+02&1.15E-05\\		
		1998&LZ Mus&9.5&4&&1.27&-8.16&4.15E+02&2.85E-06\\
		1919&V849 Oph&11.2&140&&0.82&-8.90&2.80E+04&3.54E-05\\
		1988&V2214 Oph&12.0&60&&0.99&-9.24&4.23E+04&2.41E-05\\
		1991&V2264 Oph&11.0&22&&1.10&-8.81&1.06E+04&1.64E-05\\
		1993&V2295 Oph&11.7&9&&1.19&-9.11&1.15E+04&8.83E-06\\
		1994&V2313 Oph&12.5&8&&1.20&-9.46&2.15E+04&7.44E-06\\
		2002&V2540 Oph&12.9&66&&0.96&-9.63&1.02E+05&2.37E-05\\
		1901&GK Per&12.8&6&&1.22&-9.59&2.19E+04&5.60E-06\\
		1925&RR Pic&11.2&73&&0.95&-8.90&2.05E+04&2.59E-05\\
		1991&V351 Pup&13.2&9&&1.19&-9.76&4.63E+04&7.96E-06\\
		2004&V574 Pup&10.2&12&&1.15&-8.47&3.19E+03&1.09E-05\\
		1977&HS Sge&13.5&15&&1.13&-9.89&9.20E+04&1.17E-05\\
		1936&V732  Sgr&9.6&65&&0.95&-8.21&3.41E+03&2.13E-05\\
		1977&V4021 Sgr&9.1&56&&0.98&-7.99&1.78E+03&1.82E-05\\
		1991&V4160 Sgr&12&2&&1.30&-9.24&3.25E+03&1.85E-06\\
		1992&V4169 Sgr&9.1&24&&1.09&-7.99&1.30E+03&1.33E-05\\
		1999&V4444 Sgr&13.4&5&&1.23&-9.85&3.29E+04&4.64E-06\\
		1998&V4633 Sgr&11.3&17&&1.13&-8.94&1.23E+04&1.41E-05\\
		2001&V4643 Sgr&8.3&3&&1.40&-7.64&5.32E+00&1.21E-07\\
		2001&V4739 Sgr&10.8&2&&1.30&-8.72&1.04E+03&1.97E-06\\
		2001&V4740 Sgr&11.3&18&&1.12&-8.94&1.28E+04&1.46E-05\\
		2002&V4742 Sgr&10.1&9&&1.18&-8.42&2.28E+03&8.62E-06\\
		2002&V4743 Sgr&11.8&6&&1.22&-9.16&8.89E+03&6.18E-06\\
		2003&V4745 Sgr&9.7&79&&0.92&-8.25&4.04E+03&2.28E-05\\
		2004&V5114 Sgr&12.9&9&&1.19&-9.63&3.45E+04&8.01E-06\\
		2005&V5115 Sgr&10.1&7&&1.20&-8.42&1.83E+03&6.91E-06\\
		2005&V5116 Sgr&8.4&12&&1.22&-7.69&1.97E+02&4.06E-06\\		
\tableline
\end{tabular}
\end{center}
\end{table}
\setcounter{table}{2}
\begin{table}[h]
\begin{center}
\caption{CN: WD masses and accretion rates (\textit{cont.})}
\begin{tabular}{llccccccc}   
\tableline\tableline
\multicolumn{4}{c}{Observations}&&\multicolumn{4}{c}{Modeling}\\
\cline{1-4}\cline{6-9}
&&&&&&&&\\
Year&Nova&A(mag)&$t_2$(days)&&$M_{\rm WD}(M_\odot)$&$\log\dot{M}(M_\odot$/yr)&$p_{rec}$(yr)&$m_{acc}(M_\odot)$\\
\tableline
		1992&V992 Sco&9.5&100&&0.89&-8.16&3.38E+03&2.33E-05\\
		2004&V1186 Sco&8.3&12&&1.24&-7.64&1.46E+02&3.34E-06\\
		2004&V1187 Sco&8.2&10&&1.28&-7.60&6.28E+01&1.58E-06\\
		2005&V1188 Sco&10.1&11&&1.16&-8.42&2.67E+03&1.01E-05\\
		1975&V373 Sct&12.2&47&&1.02&-9.33&4.70E+04&2.19E-05\\
		1989&V443 Sct&11.5&33&&1.06&-9.03&2.10E+04&1.97E-05\\
		1970&FH Ser&12.3&49&&1.02&-9.37&5.25E+04&2.22E-05\\
		1978&LW Ser&11.1&32&&1.06&-8.85&1.40E+04&1.96E-05\\
		1999&V382 Vel&13.8&6&&1.21&-10.02&6.13E+04&5.80E-06\\
		1968&LV Vul&10.8&20&&1.11&-8.72&8.28E+03&1.56E-05\\
		1976&NQ Vul&11.0&21&&1.10&-8.81&1.04E+04&1.60E-05\\
		1984&PW Vul&10.5&44&&1.01&-8.60&8.49E+03&2.16E-05\\
		1984&QU Vul&12.6&20&&1.12&-9.50&4.58E+04&1.43E-05\\
		1987&QV Vul&10.9&37&&1.04&-8.77&1.21E+04&2.07E-05\\
\tableline
\end{tabular}
\end{center}
\end{table}
\begin{table}[h]
\begin{center}
\caption{RN: WD masses and accretion rates from observations \citep{sch10,str10} and models}
\begin{tabular}{lcccccc}   
\tableline\tableline
\multicolumn{3}{c}{Observations}&&\multicolumn{3}{c}{Modeling}\\
\cline{1-3}\cline{5-7}
&&&&&&\\
Nova&$f$(days)&$p_{rec,obs}$(yr)&&$M_{\rm WD}(M_\odot)$&$\log\dot{M}(M_\odot$/yr)&$m_{acc}(M_\odot)$\\
\tableline
                T Pyx            &   329   &   19 (44)   &&   1.23   &   -6.95 (-7.20)  &   2.04E-06 (2.74E-06) \\
                IM Nor          &   424   &   82 (41)  &&   1.21   &   -7.32 (-7.11)   &   3.87E-06 (3.07E-06)\\
	        CI Aql           &   429   &   24   &&   1.21   &   -6.95   &   2.55E-06\\
	        V2487 Oph  &     52   &   18   &&   1.35   &   -7.39   &   7.23E-07\\
	        U Sco          &     42   &   10   &&   1.36   &   -7.29   &    5.26E-07\\
	        V394 CrA    &     65   &    30  &&    1.34  &   -7.49   &    9.72E-07\\
		T CrB          &     93   &    80  &&    1.32  &   -7.68   &    1.65E-06\\
		RS Oph       &   114   &    15  &&    1.31  &   -7.14  &    1.04E-06\\
		V745 Sco    &     19   &   21   &&    1.40  &  -7.96    &    2.32E-07\\
		V3890 Sgr  &      30  &   22   &&    1.38  &   -7.69   &     5.09E-07\\
\tableline
\end{tabular}
\end{center}
\end{table}

For RNe, we again need two observables that can be fitted by model results, so that the relations can be inverted to obtain the WD mass and the accretion rate. Since the ranges of amplitudes and decline times are not sufficiently wide to provide reliable fit functions, we use two other observables, which are available for RNe: the flash duration $f$ (i.e. the time from first brightening to final decline) and the average recurrence period $p_{rec}$, both of which we take from \citet{sch10}. We thus derive new fit formulae $f(M_{\rm WD},\dot M)$ and $p_{rec}(M_{\rm WD},\dot M)$ based on the denser grid of RNe models calculated by \cite{hil16} for a WD core temperature of 30 MKelvins, and invert them, as described there, to obtain $M_{\rm WD}(f,p_{rec})$ and $\dot M(f,p_{rec})$ from the observables of \citet{sch10}. We note that $p_{rec}$ is denoted by $D$ in \cite{hil16}. The results are presented in Table 4, where the first column lists the year of the last eruption. For IM Nor, the recurrence period is uncertain (an eruption may have been missed), while for T Pyx, the last recurrence time was significantly larger than the average; hence for these two objects we list values obtained for two possible recurrence periods.

We note that both samples on which the mass and accretion rate distributions are based---the observational
one and the theoretical one---are sparse. Hence statistically, the individual results have significant errors (we
estimate ${0.1 M_\odot}$ or so, based on deviations of assumed WD masses and accretion rates from the polynomial interpolation functions noted above). This is acceptable because the aim of this paper is to determine the nova WD mass {\it distribution}, rather than accurate individual masses.   
The few extant dynamical WD mass estimates (see below) have similar errors.

\subsection{Comparison with WD masses derived from observations}

There are only five direct, dynamical determinations of WD masses from observations in the literature. In Table 5 we compare our results with those five WD masses.

\begin{table}[ht]
\begin{center}
\caption{Comparison of WD masses (in $M_\odot$) derived from observations and models}
\smallskip
\begin{tabular}{lccl}
\tableline\tableline
{Nova} & {Model} & {Observation} & {Reference}\\
\tableline
V838 Her 1991 & $1.35$ & $1.38 \pm 0.13$ & \cite{gar18} \\
BT Mon 1939 & $0.95$ & $1.04\pm 0.06$ & \cite{smi98} \\
RN U Sco & $1.36$ & $1.55\pm 0.24$ & Thoroughgood et al. (2001)\\
V603 Aql 1918 & $1.24$ & $1.2\pm 0.2$ & Arenas et al. (2001)\\
DQ Her 1934 & $0.95$ & $0.60\pm 0.07$ & \cite{hor93} \\
\tableline
\end{tabular}
\end{center}
\end{table}
\noindent For 4 out of 5 cases - BT Mon, U Sco, V838 Her and V603 Aql - we find very good agreement between the WD masses derived directly from observations and our theoretical derivation. 
We also note that the ejected mass we calculate for BT Mon was derived from observations to be $\sim3\times10^{-5}M_{\rm WD}$ \citep{sch83} , in very good agreement with the mass derived here.             

DQ Her is a notable exception, where the most recent mass determination 0.60+/-0.07 $M_\odot$ \citep{hor93} contrasts with our theoretical determination of 0.9 $M_\odot$. The spectrographic data reduction is complex, as explained in detail by \cite{hor93}. Two earlier estimates claim masses of $1.09M_\odot$ \citep{rob76} and $1.0\pm0.1M_\odot$ \citep{hut79}. \cite{hor93} themselves report on a calculation with slightly less stringent assumptions that results in a higher WD mass, $0.68\pm0.1M_\odot$. DQ Her is unusual as its WD is a very fast rotator (with a spin period of 71~s). Our own derivation may be an overestimate; when the critical pressure required for ignition is reached, the accreted column mass at the pole is lower than that at the equator. This means that the outburst will start declining earlier, implying a WD mass larger than it really is. In addition, DQ Her showed a deep dip in its light curve beginning 100 days after maximum.

\begin{figure}[h]
\includegraphics[scale=0.70]{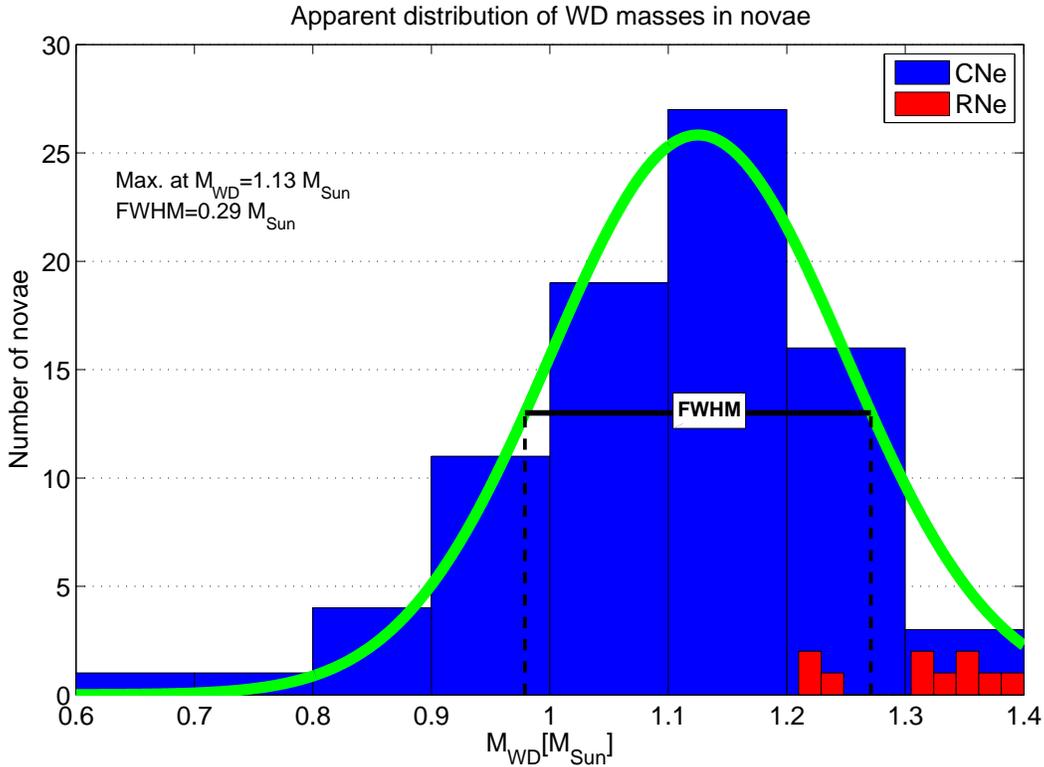}
\caption{The frequency-averaged distribution of WD masses for 82 CN and 10 RNe. See text for details.}
\end{figure}

\subsection{Comparison with WD masses derived by light-curve fitting}

The data bases on which the present study is based are either purely observational or entirely theoretical. By combining them, as explained above, we were able to determine the WD masses of observed novae. A completely different procedure that combines theory and observations in order to derive nova WD masses was proposed by \cite{kat94}, hereafter KH94, and has been used by them for individual objects ever since. As the KH94 mass estimates are derived with an entirely different methodology from that of this paper, it is interesting to compare them to the masses derived here. The KH94 method is based on fitting an optically thick wind solution, following the theory developed by \cite{rug79}, to the observed multi-band light curve corresponding to the mass-loss and decline phase of a nova. The parameters assumed in KH94 are the WD mass, the initial envelope mass, and its composition, disregarding prior evolution and the WD structure below the envelope. A series of steady state models with decreasing envelope mass is used to mimic the progression in time. The parameters are varied until agreement with the observed light curve is achieved. The WD masses obtained by the two methods of combining theory and observation are compared in Table 6. 

\begin{table}[ht]
\begin{center}
\caption{Comparison of WD masses with those obtained by light-curve fitting}
\smallskip
\begin{tabular}{llccc}
\tableline\tableline
Paper&Object&$M_{\rm WD}$ from K\&H&$M_{\rm WD}$ this paper & $t_2$ days$^{(1)}$\\ 
\tableline
\cite{hac16}&V1668 Cyg$^{(a)}$&0.98&1.16& 11 \\
		&V1974 Cyg$^{(b)}$&0.98&1.12& 19 \\
		&QU Vul&0.96&1.04& 20 \\
		&V351 Pup&1.00&1.19& 9 \\
		&V382 Vel$^{(c)}$&1.23&1.21& 6 \\		
		&V693 CrA&1.15&1.15&10 \\
		\tableline
\cite{hac15}&PW Vul&0.83&1.01& 44 \\
		&V705 Cas&0.78&1.05& 33 \\
		&RR Pic&0.5-0.6&0.95& 73 \\
		&HR Del&0.5-0.55&0.84& 167 \\
		&V723 Cas&0.5-0.55&0.77& 263\\
		\tableline
\cite{hac14}&V1500 Cyg$^{(d)}$&1.20&1.09& 2 \\
		&GK Per$^{(e)}$&1.05&1.22& 6 \\
		&V603 Aql&1.20&1.24& 5 \\
		\tableline
\cite{hac10}&V382 Vel 1999$^{(c)}$&1.23&1.21& 6 \\
		&V4743 Sgr 2002&1.15&1.22& 6 \\
		&V1494 Aql 1999&1.13/1.06/0.92&1.20& 8 \\
		&V2467 Cyg 2007&1.11/1.04/0.90&1.20& 8 \\
		&V5116 Sgr 2005&1.07/1.0/0.85&1.22& 12 \\
		&V574 Pup 2004&1.05&1.15& 12 \\
		&V1974 Cyg 1992$^{(b)}$&1.05&1.12& 19 \\
		&V1668 Cyg 1978$^{(a)}$&0.95&1.16& 11 \\
		&V1500 Cyg 1975$^{(d)}$&1.15&1.09& 2 \\
		\tableline
\cite{hac09}&V2491 Cyg&1.32/1.3/1.27&1.26& 4 \\
		&V1493 Aql&1.2/1.15/1.1&1.18& 9 \\
		&V2362 Cyg&0.75/0.7/0.65&1.19& 9 \\
		\tableline
\cite{kat09}&V838 Her&1.35$\pm$0.02&1.35& 1 \\
		\tableline
\cite{hac07}&GK Per$^{(e)}$&1.15$\pm$0.05&1.22& 6 \\
		&V5115 Sgr&1.20$\pm$0.1&1.20& 7 \\
		&V5116 Sgr&0.90$\pm$0.1&1.22& 12 \\
		\tableline
\cite{hac03}&Cl Aql&1.2$\pm$0.05&1.21& 25 \\
		\tableline
\cite{hac01}&T CrB&1.37&1.32& 4 \\
		&RS Oph&1.35/1.377&1.31& 7 \\
		&V745 Sco&1.35&1.4&\\
		&V3890 Sgr&1.35&1.38& 6 \\
		&U Sco&1.37&1.36& 1 \\
		&V394 CrA&1.37&1.34&\\

\tableline
\end{tabular}
\newline{ $^{(1)}$ from \cite{str10}, \qquad $^{(a),(b),(c),(d),(e)}$ multiple entries}
\end{center}
\end{table}

Despite the inherent approximations and simplifications of each method, the agreement between them is acceptable for now. Future direct observational WD mass measurements will be essential to determine which method of WD mass derivation is more accurate.

\subsection{Apparent distribution of WD masses and accretion rates}

The distributions of WD masses and accretion rates that we derive are listed in Tables 1 and 2 and are shown in Figs. 2 and 3. The average nova WD mass is $1.13 M_\odot$ and the average accretion rate $1.25\times10^{-9}M_\odot$/yr. We note that the average mass is in excellent agreement with the theoretical estimate of \cite{rit91}, who predict the mean WD mass, weighted by nova frequency (that is, the value determined directly by observations), to be found between 1.04 and 1.24 $M_\odot$. The peak of the accretion rate distribution corresponds to $\dot M=10^{-9}M_\odot$/yr, which is in excellent agreement with \cite{tow05}, who argue, based on the distribution of orbital periods, that about 50\% of CN occur in binaries accreting at $\dot M\approx10^{-9}M_\odot$/yr.

\begin{figure}[h]
\includegraphics[scale=0.70]{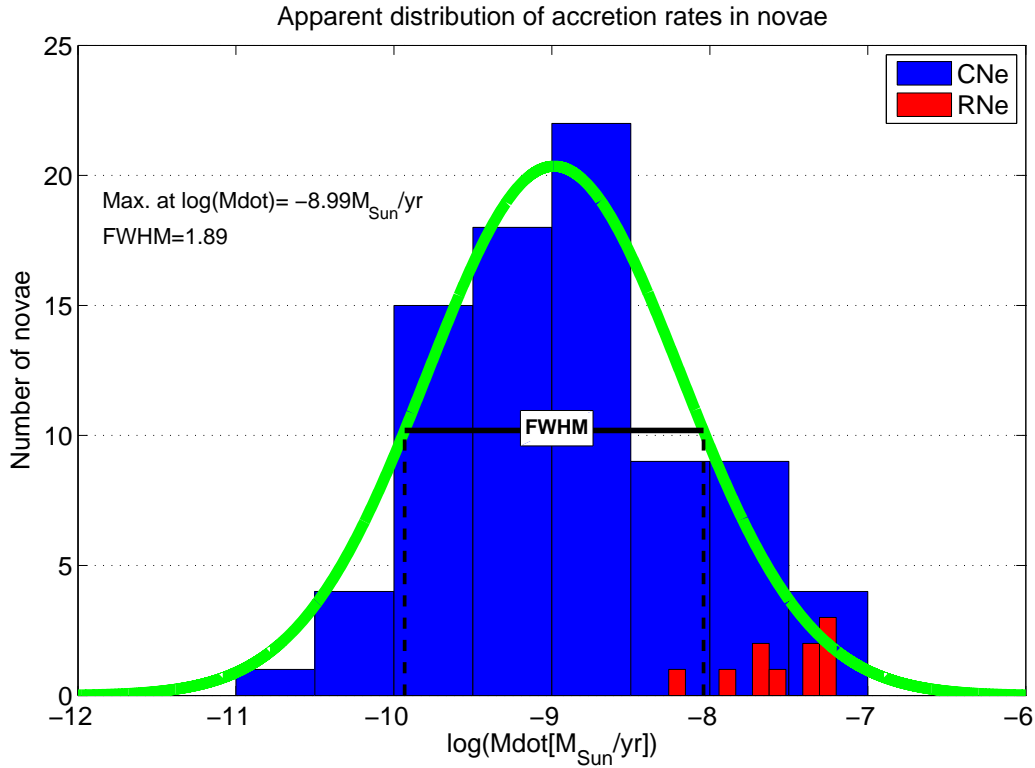}
\caption{The frequency-averaged distribution of accretion rates for 82 CN and 10 RNe. See text for details.}
\end{figure}

In Fig. 4 we show the recurrence time distribution of our sample of 82 CN and 10 RNe. Recurrence times are determined by the amount of accreted mass required to trigger an outburst, which depends strongly on the WD mass \citep{pri95}; they are inversely proportional to the accretion rate.
\begin{figure}[h]
\includegraphics[scale=0.70]{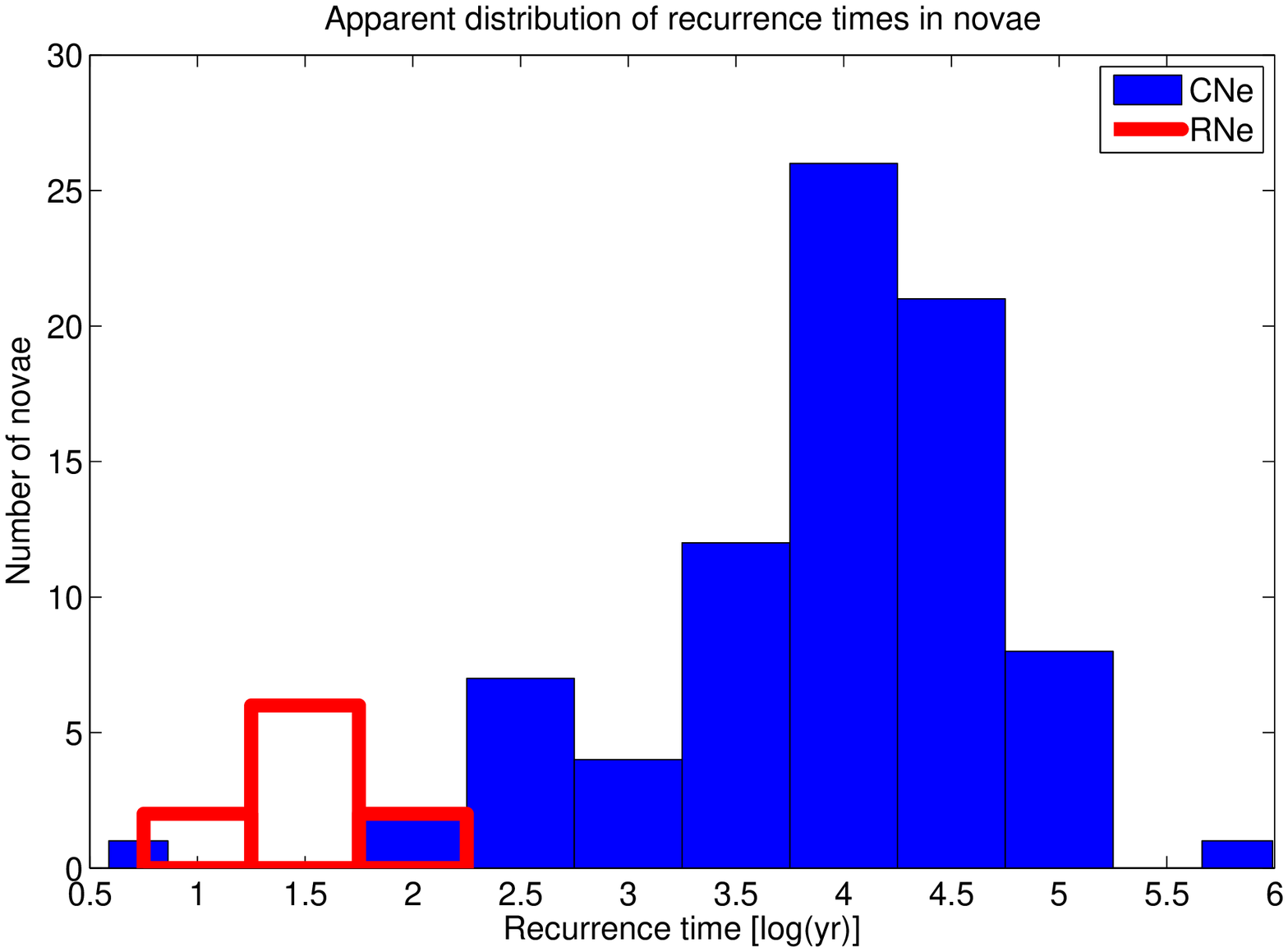}
\caption{The distribution of recurrence times for 82 CN and 10 RNe. See text for details.}
\end{figure}
The distributions of WD masses and accretion rates obtained for the observed sample do not reflect the true distributions within nova systems, since they are strongly biased by the outburst recurrence times. 

\section{Distribution of WD masses and accretion rates in novae}

In this section we derive the true distributions of WD masses and accretion rates in nova systems, based on the sample of observed novae in Tables 1 and 2. The sample 
is assumed to be sufficiently large and randomly sampled that even quite rarely erupting novae are included. We note that, if they exist, extremely low mass transfer rate systems (i.e. ``hibernating" cataclysmic binaries \citep{sha86}) with $\dot M \leq$  $10^{-11}M_\odot$/yr will be missed; only one such system - V1500 Cyg - is identified in Table 1.

Let $x$ denote the WD mass and $P_1(x)$ the probability that the WD mass in a nova systems is $x$. By contrast, the probability that a WD from the observed sample has mass $x$ will be denoted $P_{1,obs}(x)$. Let $y$ denote the accretion rate (on logarithmic scale) and $P_2(y)$ -- the probability that the accretion rate in a nova system is $y$. Similarly, the probability that a nova in the observed sample has accretion rate $y$ (on logarithmic scale) will be denoted by $P_{2,obs}(y)$. The variables $x$ and $y$ may assume values within known intervals $[x_{min},x_{max}]$ and $[y_{min},y_{max}]$. 

For the sample of $N$ observed novae and the distribution of masses and accretion rates within them, we define the corresponding  probabilities $P_{1,obs}(x)$ and $P_{2,obs}(y)$ as follows: if $N_{1,obs}(x)dx$ denotes the number of systems with masses in the range $[x,x+dx]$ and $N_{2,obs}(y)dy$ -- the number of systems with accretion rates in the range $[y,y+dy]$, then
\begin{eqnarray}
dN_{1,obs}(x)&=&P_{1,obs}(x)dx\cr
dN_{2,obs}(y)&=&P_{2,obs}(y)dy\ 
\end{eqnarray}
and obviously,
\begin{equation}
\int_{x_{min}}^{x_{max}} P_{1,obs}(x)dx =  \int_{y_{min}}^{y_{max}} P_{2,obs}(y)dy  = N,
\end{equation}
the total number of objects in the observed sample.

The difference between the actual mass and accretion rate distributions and the observed ones is due to different recurrence periods $p_{rec}$ of nova outbursts, where the recurrent period is a function of both the WD mass and the accretion rate. This functional dependence, which we denote by $p_{rec}=f(x,y)$, is derived theoretically from the grid of models.

Now, the probability of observing a system with a WD mass $x$ and an accretion rate $y$ per unit time is proportional to the occurrence frequency $p_{rec}^{-1}$, hence the probabilities of observing systems of given mass or given accretion rate are given by 
\begin{eqnarray}\label{eq:prob}
P_{1,obs}(x)& = &P_1(x)\int_{y_{min}}^{y_{max}} \frac{1}{f(x,y)}P_{2,obs}(y)dy\cr
&&\cr
P_{2,obs}(y)& = &P_2(y)\int_{x_{min}}^{x_{max}} \frac{1}{f(x,y)}P_{1,obs}(x)dx\ .
\end{eqnarray}
where the observed distributions are known. We can solve Eq.\ref{eq:prob} to obtain $P_1(x)$ and $P_2(y)$, either by assuming Gaussians for $P_{1,obs}(x)$ and $P_{2,obs}(y)$ or by discretization. We thus obtain relative distributions (that we normalize), namely, the shifts of the observed distribution with respect to the actual ones. Obviously, if the recurrence time is the same for all objects, the probability distributions will be equal.
We apply this formalism to the sample of galactic novae by binning the WD mass range into $n_x$ bins ${x_1, x_2, \dots , x_{n_x}}$ and the accretion rates into $n_y$ bins ${y_1, y_2, \dots , y_{n_y}}$. To each observed nova we assign a $x_i$ value according to its corresponding mass bin and a $y_j$ value, according to its corresponding accretion rate bin. We then calculate the function $f(x_i,y_j)$ according to a fit formula derived from the grid of models (in a similar way as we computed WD masses and accretion rates).

The number of objects in each mass bin $x_i$ is denoted $N_{x,i}$ and the number of objects in each mass accretion bin, $N_{y,j}$, and hence 
\begin{equation}
P_{1,obs}(x_i)=N_{x,i}/N
\end{equation}
and similarly,
\begin{equation}
P_{2,obs}(y_j)=N_{y,j}/N\ .
\end{equation}
Now, for each $x_i$ and $y_j$, the integrals on the RHS of Eq.\ref{eq:prob} are replaced by the following sums, respectively:
\begin{equation}
S_x(x_i)=\sum_{j=1}^{n_y}\frac{P_{2,obs}(y_j)}{f(x_i,y_j)} \qquad {\rm and} \qquad S_y(y_j)=\sum_{i=1}^{n_x}\frac{P_{1,obs}(x_i)}{f(x_i,y_j)}\ ,
\end{equation}
where the summation includes all objects that fall into the WD mass bin $x_i$ in the first case, and all objects that fall into the accretion rate bin $y_j$, in the second. This yields the original distributions $P_1(x_i)=P_{1,obs}(x_i)/S_x(x_i)$ and $P_2(y_j)=P_{2,obs}(y_j)/S_y(y_j)$, which we normalize, so that 
\begin{equation}
\sum_{i=1}^{n_x}P_1(x_i)=1 \qquad {\rm and} \qquad \sum_{j=1}^{n_y}P_2(y_j)=1\ .
\end{equation} 
\begin{figure}[h]
\includegraphics[scale=0.70]{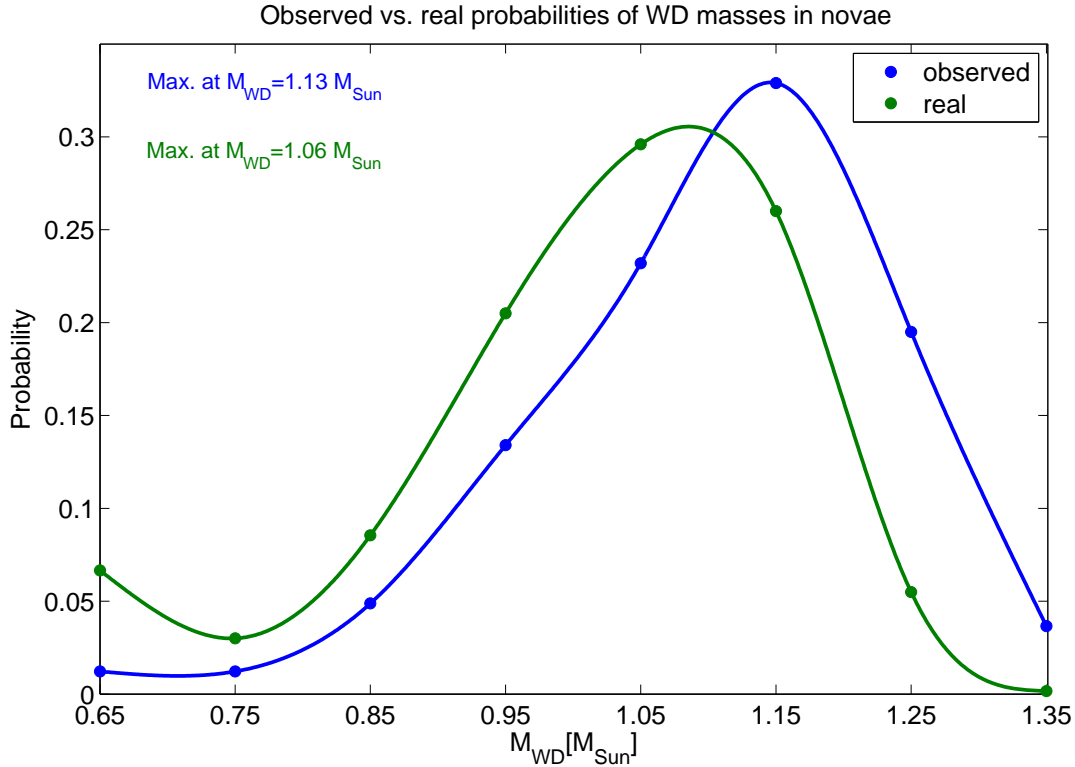}
\caption{The real distribution of WD masses in nova systems. See text for details.}
\end{figure}

\begin{figure}[h]
\includegraphics[scale=0.70]{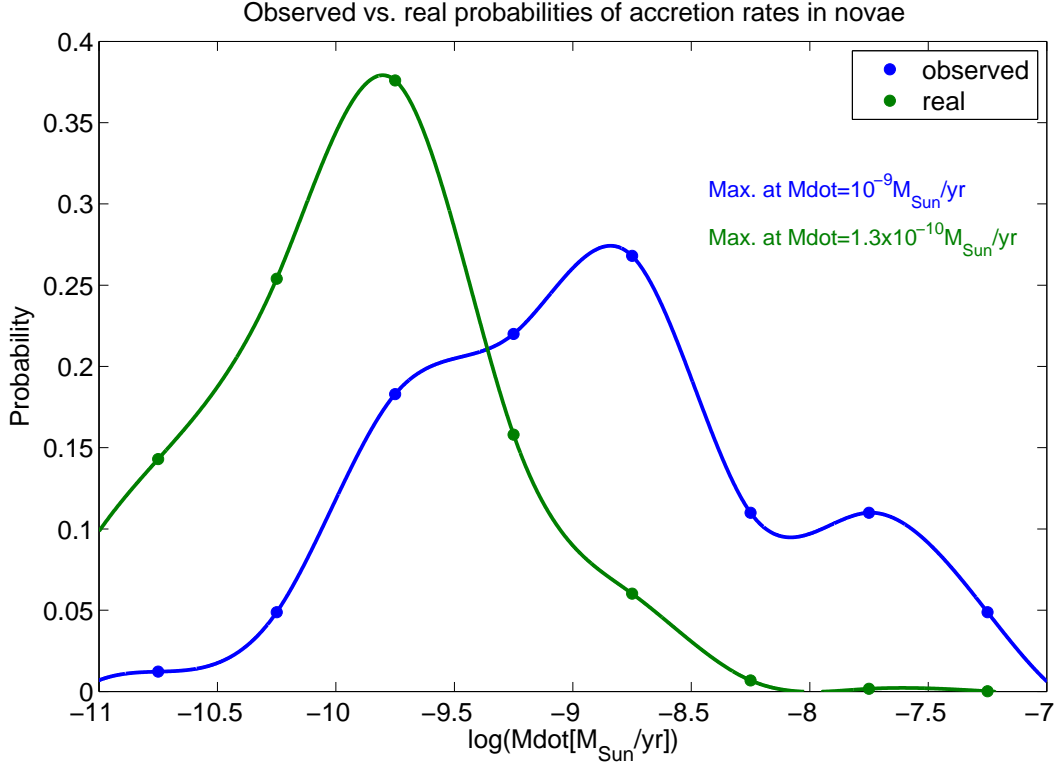}
\caption{The real distribution of accretion rates in nova systems. See text for details.}
\end{figure}

The results for the WD mass distribution, showing the shift with respect to the observed distribution are shown in Fig. 5. The accretion rate distribution, where the shift amounts to one order of magnitude, is shown in Fig. 6. The most probable WD mass is $1.13M_\odot$ while the average WD mass of the true distribution is $1.06 M_\odot$. The true average accretion rate $1.3\times10^{-10}M_\odot$/yr.

\section{Discussion and Conclusions}

The frequency-averaged masses of WDs in CN and RNe have long been predicted to be much larger than those of field WD.
We have used our extensive set of nova models to determine the relationships between WD mass, time to decline by 2 magnitudes, 
and outburst amplitude. These relationships have been used to deduce the masses of the WD in 82 Galactic CN.

We find that while the average CN mass of the Galactic novae is 1.13 $M_\odot$, the frequency-averaged (true) CN mass is 1.06 $M_\odot$, in very good agreement with population synthesis predictions. The mean mass of 10 recurrent novae is significantly larger, 1.31 $M_\odot$. 

The difference in WD mass between the frequency-averaged mass and the true average mass is modest - just 0.07 $M_\odot$, so that the sharp discrepancy between field WD and nova WD masses remains in place, and important to understand. The true mass transfer rate, however, {\it is a full order of magnitude lower than the frequency-averaged rate}, (i.e. that corresponding to direct observation). The most frequent outbursts, detected most frequently, should occur in the highest  $\dot M$ systems. Here, for the first time, we are able to quantify the size of that observational bias.

We find that the longterm, true mass accretion rate of CN is remarkably low: $1.3\times10^{-10}M_\odot$/yr. All cataclysmic binaries transferring matter at
this low rate experience dwarf nova eruptions. This is supportive of the view that mass transfer rates decline strongly in the centuries after nova eruptions, so that old novae undergo a metamorphosis to become dwarf novae for some or much of the millennia between nova eruptions. 
The four oldest old novae are all now observed to be dwarf novae \citep{sha07,sha12,mis16, shb17} surrounded by shells of ejected matter.

We determine the average mass accretion rates of the 10 known Galactic RN to be in the range $10^{-7}$ - $10^{-8}$ $ M_\odot$/yr (with a mean of $4\times10^{-8}$ $ M_\odot$/yr). 

Finally, we have determined the recurrence time distribution of Galactic novae. In addition to RN with recurrence times shorter than a century, 
we find a few CN which must erupt every few hundred years. However, the vast majority of CN recur on timescales of tens of millennia or longer.
We note that in the sample of CN, that is novae for which only one outburst has been recorded, there are two objects with calculated recurrent times falling within the range of RNe (less than 100~yr): V1187 Sco and V4643 Sgr, both relatively recent and very fast novae. We predict that these are in fact RNe and the latter, at least, may erupt again in the near future.

MMS gratefully acknowledges the support of the late Ethel Lipsitz and Hilary Lipsitz, longtime friends of the AMNH Astrophysics department. 
He thanks J. Mikolajewska for very helpful suggestions concerning WD masses in recurrent novae.

\end{document}